\documentclass{13akws-procs9x6}

\def\etal{{\it et al.}}

\newcommand{\bea}{\begin{eqnarray}}
\newcommand{\eea}{\end{eqnarray}}

\begin{document}

\title{COMMENTS ON LORENTZ AND CPT VIOLATION}

\author{V.\ ALAN KOSTELECK\'Y}

\address{Physics Department, Indiana University\\
Bloomington, IN 47405, USA}

\begin{abstract}
This contribution to the CPT'13 meeting
briefly introduces Lorentz and CPT violation
and outlines two recent developments in the field.
\end{abstract}

\bodymatter

\section{Introduction}

The idea that small observable violations of Lorentz symmetry 
could provide experimental access to Planck-scale effects
continues to draw attention across several subfields of physics.
In the three years since the previous meeting in this series,
considerable progress has been made 
on both experimental and theoretical fronts.
This contribution to the CPT'13 proceedings
contains a brief introduction,
followed by comments on two topics of recent interest:
nonminimal fermion couplings
and Riemann-Finsler geometry.

\section{Basics}

A satisfactory theoretical description of Lorentz violation
must incorporate coordinate independence, realism, and generality.
A powerful approach uses effective field theory,
starting with General Relativity coupled to the Standard Model
and adding to the Lagrange density all observer-invariant terms 
that contain Lorentz-violating operators
combined with controlling coefficients.
This yields the comprehensive realistic effective field theory 
for Lorentz violation
called the Standard-Model Extension (SME)
\cite{ck,akgrav}.
The SME also describes general CPT violation,
which in the context of realistic effective field theory
is accompanied by Lorentz violation 
\cite{owg}.
The full SME contains operators of arbitrary mass dimension $d$,
while the minimal SME restricts attention 
to operators of renormalizable dimension $d \leq 4$.

Observable signals of Lorentz violation
are governed by the SME coefficients.
Experiments typically search for particle interactions 
with background coefficient values,
which can produce effects
dependent on the particle velocity, spin, flavor, and couplings.
Many investigations of this type have been performed
\cite{tables},
achieving impressive sensitivities 
that in some cases exceed expectations
for suppressed Planck-scale effects. 
If the SME coefficients are produced by 
spontaneous Lorentz violation 
\cite{ksp},
as is necessary when gravity is based on Riemann geometry
\cite{akgrav},
then they are dynamical quantities
that must incorporate massless Nambu-Goldstone modes
\cite{lvng}.
These modes have numerous interpretations,
including serving as an alternative origin 
for the photon in Einstein-Maxwell theory
\cite{lvng}
and the graviton in General Relativity 
\cite{nggrav}, 
or representing new spin-dependent 
\cite{ahclt}
or spin-independent 
\cite{ngnospin}
forces,
among other possibilities.
Massive modes can also appear
\cite{lvmm}.

\section{Nonminimal fermion sector}

In the nonminimal sector of the SME,
the number of Lorentz-violating operators 
grows rapidly with the mass dimension $d$.
Systematically cataloguing and characterizing the possibilities
is therefore indispensible 
in the search for Lorentz violation.
In the CPT'10 proceedings,
I outlined some features appearing in  
the treatment of quadratic operators 
of arbitrary $d$ in the photon Lagrange density
\cite{km09}.
In the intervening three-year period,
investigations of the quadratic fermion sector
for arbitrary $d$ have also been performed.
The Lagrange density for propagation and mixing of any number of fermions 
has been developed and applied to describe 
general Lorentz violation in the neutrino sector 
\cite{km12}.
More recently,
quadratic operators of arbitrary $d$
have been studied for a massive Dirac fermion
\cite{km13}.

Many nonminimal operators generate effects
that are in principle observable,
and each such operator generates a distinct experimental signal.
For quadratic operators,
which characterize particle propagation
and phase-space features of particle decays,
the Lorentz-violating effects can include 
direction dependence (anisotropy),
wave-packet deformation (dispersion),
and mode splitting (birefringence).
In the neutrino sector,
for example,
some operators control flavor-dependent effects 
in neutrino and antineutrino mixing,
producing novel energy and direction dependences
that involve both Dirac- and Majorana-type couplings.
Others govern species-independent effects,
which can differ for neutrinos and antineutrinos
and can produce propagation times
varying with energy and direction,
in some cases exceeding that of light. 
A few operators produce `countershaded' effects\cite{akjt} 
that cannot be detected via oscillations or propagation
but change interaction properties
in processes such as beta decay
\cite{dkl}. 

Analogous effects appear in the description 
of a massive Dirac fermion in the presence of Lorentz violation,
for which the exact dispersion relation for arbitrary $d$ 
is known in closed and compact form
\cite{km13}.
For example,
the fermion group velocity is anisotropic and dispersive,
while the fermion spin exhibits a Larmor-like precession 
caused by birefringent operators. 
Using field redefinitions to investigate observability
reveals that many operators of dimension $d$ produce no effects
or are physically indistinguishable from others
of dimensions $d$ or $d\pm1$.
Nonetheless,
the number of observable coefficients grows as the cube of $d$.
To date,
almost all the nonminimal coefficient space 
for fermions is experimentally untouched.
This offers an open arena for further exploration
with a significant potential for discovery.

\section{Riemann-Finsler geometry}

The surprising `no-go' result 
that the conventional Riemann geometry
of General Relativity and its extension to Riemann-Cartan geometry 
are both incompatible with explicit Lorentz violation
raises the questions of whether an alternative geometry is involved
and, if so, whether a corresponding gravitational theory exists.
The obstruction to explicit Lorentz violation,
which disappears for the spontaneous case,
can be traced to the generic incompatibility of the Bianchi identities
with the external prescription of coefficients for Lorentz violation. 
It is therefore reasonable to conjecture
that a natural geometrical setting would include
metric distances depending locally on the coefficients
in addition to the Riemann metric.
\cite{akgrav}

Support for this conjecture has recently emerged
with the discovery that a fermion 
experiencing explicit Lorentz violation
tracks a geodesic in a pseudo-Riemann-Finsler geometry
rather than a conventional geodesic in pseudo-Riemann spacetime
\cite{ak11}.
Riemann-Finsler geometry is a well-established mathematical field
with numerous physical applications
(see, e.g., Ref.\ \refcite{chernshen}),
such as the famous Zermelo navigation problem 
of obtaining the minimum-time path 
for a ship in the presence of ocean currents.
A large class of Riemann-Finsler geometries
is determined by the geodesic motion of particles in the SME
\cite{ak11,finsler}.
Among these are the canonical Randers geometry,
which is related to the 1-form SME coefficient $a_\mu$,
and numerous novel geometries of simplicity 
comparable to the Randers case.
One example of the latter is $b$ space,
a calculable Riemann-Finsler geometry that also is based on a 1-form 
and has Finsler structure complementary to that of Randers space.
Physically,
this geometry underlies the geodesic motion of a fermion 
in curved spacetime in the presence of chiral CPT-odd Lorentz violation.
All the SME-inspired geometries
exhibit mathematically interesting features
connected to physical properties.
For instance,
when the SME coefficients are covariantly constant,
the trajectories are conventional Riemann geodesics
and special Riemann-Finsler geometries known as Berwald spaces result.
Many open challenges remain in this area,
ranging from more technical questions such as resolving singularities 
or classifying geometries
to physical issues such as generalizing Zermelo navigation
or uncovering implications for the SME.
The prospects appear promising for further insights
to emerge from this geometrical approach to Lorentz violation.

\section*{Acknowledgments}

This work was supported in part 
by U.S.\ D.o.E.\ grant DE-FG02-13ER42002
and by the Indiana University Center for Spacetime Symmetries.

\end{document}